\documentstyle[aps,twocolumn,epsfig]{revtex}
\begin{document}
%\draft

\title{Stochastic versus
dynamic approach to L\'{e}vy statistics in the presence of an external
perturbation}

\author{Mario Annunziato$^{1,*}$,
%\footnote{E-mail: annunzia@mailbox.difi.unipi.it}
Paolo Grigolini$^{1,2,3,\dag}$
%\footnote{E-mail: grigo@jove.acs.unt.edu}
}
\address{$^{1}$Dipartimento di Fisica dell'Universit\`{a} di Pisa,
Piazza Torricelli 2, 56127 Pisa, Italy }
\address{$^{2}$Center for Nonlinear Science, University of North
Texas, P.O. Box 305370, Denton, Texas 76203 }
\address{$^{3}$Istituto di Biofisica del Consiglio Nazionale delle
Ricerche, Via San Lorenzo 26, 56127 Pisa, Italy }
\address{$^*$E-mail: annunzia@mailbox.difi.unipi.it,
 $^\dag$E-mail: grigo@jove.acs.unt.edu}
\date{\today}

\maketitle

\begin{abstract}
We study the influence of a dissipation process on
diffusion dynamics triggered by slow fluctuations.
We study both strong- and weak-friction regime. When the latter
regime applies, the system is attracted by the basin of 
either Gauss or L\'evy statistics according to whether the 
fluctuation correlation function is integrable or not.
We analyze with a numerical calculation the border 
between the two basins of attraction.
\end{abstract}


\section{Introduction}
There is a growing interest in literature\cite{nature,physicstoday}
 on the physical
manifestation of L\'{e}vy diffusion. This interesting subject
can be dealt with in two distinct ways. The first rests on the
assumption that there exist in nature L\'{e}vy fluctuations, namely
stochastic processes of the L\'{e}vy stable form\cite{levy,doob}.
We refer to this approach as \emph{stochastic}.
There are many examples of this attitude, and we limit ourselves
to quoting a sample of the early work based on this
view \cite{montroll,seshadri,west}. One of the most recent examples of
this kind of approach is given by the work of Ref.\cite{metzler}.

The second way of dealing with L\'{e}vy fluctuations is based on the
assumption that these processes have a Hamiltonian foundation. We
refer to this kind of approach as \emph{dynamic}.  An
outstanding champion of this view is Zaslavsky and the interested
reader is referred to his recent book\cite{zaslavsky} for a
transparent illustration of this perspective. According to
Zaslavsky's analysis, the Hamiltonian dynamics  yielding L\'{e}vy
diffusion is characterized by specific weak-chaos properties. For
calculation purposes these properties can be mimicked by either maps
\cite{stationary,firstderivation,trefan} of the same kind as those
used to study intermittency \cite{geisel} or a suitable non-linear
transformation of a random noise generator \cite{allegro}.
Therefore we shall refer ourselves to all these
papers\cite{stationary,firstderivation,trefan,allegro} as examples
of a dynamic approach to L\'{e}vy diffusion not to speak of those papers
explicitly using Hamiltonian systems as diffusion
generators\cite{eggcrate,hamiltonianklafter}.

As far as the case of free diffusion is concerned, the dynamic and
stochastic approach yield almost indistinguishable results.
The main purpose of this letter is to show that when the interesting
case of a perturbation is considered, the dynamic approach and the
stochastic approach can lead to totally different predictions. A
reconciliation of the two approaches can only be obtained in the
limiting case of external perturbations of extremely weak intensity.

\section{Stochastic approach}

We consider a free motion of a point particle
subject to a Markovian stochastic noise $\eta(t)$.
This means that for any infinitesimal time 
interval $dt$ there is a momentum change per unit of mass given by:
\begin{equation}
dv(t) = d\eta(t)\mbox{ ,}
\label{free}
\end{equation}
where $v(t)$ is the velocity of the particle.
In the treatment of this letter the forcing term will 
be either a Wiener or a L\'{e}vy process. This means that the 
probability density for the fluctuation $\eta(t)$ to make a jump by the 
intensity $\eta$ in the time interval  $t$, $P(\eta, t)$ , defined through 
the inverse Fourier transform of its characteristic function, reads:
\begin{equation}
{P}(\eta,t)=\frac{1}{2\pi}\int_{-\infty}^{+\infty}
e^{-ik\,\eta}\, e^{-\frac{1}{2}\sigma^2\,k^2\, t} dk
\label{Wienerproc}
\end{equation}
in the case of a Wiener process, and:
\begin{equation}
{P}(\eta,t)=\frac{1}{2\pi}\int_{-\infty}^{+\infty}
e^{-ik\,\eta}\, e^{-b\,|k|^\alpha t} dk
\label{Levyproc}
\end{equation}
in the case of a symmetric L\'{e}vy process.
The L\'evy process is characterized by a positive parameter $b$,
determining the width of the distribution and by the coefficient
$\alpha$ determining the specific L\'evy statistics of the 
system. 

As discussed in Ref. \cite{allegro}, the dynamic approach to L\'{e}vy 
processes has to be restricted to the range: $1 <\alpha <2$, while, 
as well known \cite{montroll,gnekol}, the full range of L\'{e}vy processes 
is: $0 <\alpha <2$. 
Thus, to make a comparison between the stochastic and the dynamic
approach we are forced to restrain our analysis to the restricted
interval of $\alpha$ values: $1<\alpha <2$. In this interval we do
not have available any known analytical expression for the density
distribution $P(\eta,t)$. However, we  know \cite{gnekol} that the 
distribution is bell-shaped and that in the asymptotic limit of large 
$\eta$ is characterized by inverse power law tails, proportional to  
$t/|\eta|^{(1 + \alpha)}$. As far as the width of the distribution 
is concerned, it cannot be defined by the variance, which is 
infinite. In a loose sense the width of the distribution is 
determined by the parameter $b$. It is evident that Eq.(\ref{free}) makes 
the distribution $P(v,t)$ identical to $P(\eta,t)$

Let us consider the perturbation of the free diffusion generated by 
Eq. (\ref{free}) by means of:
\begin{equation}
dv(t) = K(v)\,dt + d\eta(t) \mbox{ ,}
\label{mainequation}
\end{equation}
where $K(v)$ is a generic perturbation. 
For simplicity we restrict our analysis to the case of linear damping,
thereby replacing $K(v)$ in Eq.(\ref{mainequation}) with $-\lambda v(t)$:
\begin{equation}
dv(t) = -\lambda v(t)\, dt + d\eta(t) \mbox{ .}
\label{damped}
\end{equation}
When the fluctuation $d\eta(t)$ is the Wiener process of Eq.
(\ref{Wienerproc}), Eq.(\ref{damped}) 
becomes the ordinary Ornstein-Uhlenbeck process, which yields the 
Fokker-Planck equation and with it the equilibrium Gaussian density, 
whose Fourier transform is  $e^{-\frac{\sigma^2_\lambda}{2}k^2}$,
with variance:
\begin{equation}
\sigma_\lambda^2=\frac{\sigma^2}{2\,\lambda} \mbox{ .}
\label{sig_ou}
\end{equation} 
If $\eta(t)$ is the L\'evy process of Eq.(\ref{Levyproc}),
the Fourier transform of the equilibrium distribution
is $e^{- b_\lambda |k|^\alpha}$, 
with the same exponent $\alpha$ as that of free diffusion.
The parameter $b_{\lambda}$ (see \cite{west,metzler}) is defined by:
\begin{equation}
b_{\lambda} = \frac{b}{\alpha\,\lambda} \mbox{ .}
\label{b_sw}
\end{equation}
We note that $b_\lambda$ and $\sigma^2_\lambda$
share the same structure.
In both cases the effect of the friction term is that of 
quenching the free diffusion so as to generate a time 
independent, or equilibrium, distribution.
This means that $\lim_{t \rightarrow \infty} P(v,t) 
= {\cal P}(v)$.

\section{Dynamic approach}
Both the Wiener and L\'evy noise illustrated in Section II are 
mathematical abstractions with a limit of validity.
The discrepancy between this mathematical abstraction and physical 
reality can become significant as a result of an external perturbation
forcing the dynamics of the system to produce physical effects
stemming from the time scale where the mathematical idealization 
departs from physical reality. Here we focus our attention on the case
where the physical reality is assumed to be properly described
by the dichotomous fluctuation $\xi$ used in earlier work for a dynamic
derivation of L\'evy processes\cite{allegro}. 

We assign to this 
dichotomous variable the values $W$ and $-W$. According to Ref. \cite{allegro}
the statistics of the diffusion process generated by this variable are 
determined by both its dichotomous nature and its correlation function 
$\Phi_{\xi}(t)$. We assign to this correlation function the form:
\begin{equation}
\Phi_{\xi}(t) = \frac{(\beta\,T)^\beta}{(\beta\,T + t)^\beta}
\;\;\;\;\;\;\mbox{, $\beta > 0$.}
\label{correlationfunction}
\end{equation}
According to the theoretical analysis of Ref.\cite{allegro,geisel2}, the 
dichotomous nature of this fluctuation makes especially relevant the 
physical meaning of the function $\psi_{\xi}(t)$ defined by:
\begin{equation}
\psi_{\xi}(t) \equiv T \frac{d^{2}}{dt^{2}} \Phi_{\xi}(t) 
= \frac{(\beta + 1)(\beta T)^{\beta + 1}}{(\beta T + t)^{\beta + 2}}\mbox{ .}
\label{t_sogg}
\end{equation}
This function is the distribution of sojourn times in one of the two 
equiprobable states of the dichotomous variable $\xi$. 
The parameter $T$ denotes the mean waiting sojourn time.

It is convenient to illustrate some aspects of the free diffusion 
process generated by this fluctuation, namely, the case described by:

\begin{equation}
dv(t) = \xi(t)\,dt \mbox{ .}
\label{dicotfree}
\end{equation}
A realization of this process of free diffusion is:
\begin{equation}
v(t)=\xi_{n}\,(t-t_{n-1}^{(0)}) + \sum_{k=0}^{n-1} \xi_k \tau_k \mbox{ ,}
\label{realizdicot}
\end{equation}
$$
\mbox{where} \;\;\;
 t_{n-1}^{(0)}<t<t_{n-1}^{(0)}+\tau_n \;\;\; \mbox{ and} \;\;\;
 t_{n-1}^{(0)}=\sum_{k=0}^{n-1}\tau_k \mbox{ .}
$$
Here the $\tau_k 's$ denote the time durations of the sojourns
in the accelerating states occurring prior to time $t$,
and the $\xi_k 's$ denote either the value $W$ or the value
$-W$ of the  dichotomous fluctuation.
The distribution of the sojourn times is given by Eq. (\ref{t_sogg}). 
Thus we easily derive the distribution of the velocity jumps
$\Delta v_k = \xi_k \tau_k$, which turns out to be:
\begin{equation}
\psi(\Delta v)=\frac{(\beta+1)(\beta\,T)^{\beta+1}
}{2\,W\,(\beta\,T+|\Delta v|/W)^{\beta+2}} \mbox{ .}
\label{dist_dicot}
\end{equation}

This distribution is of central importance to predict the statistics 
of the resulting diffusion process.
If $\beta > 1$ the second moment of this distribution 
is finite. Thus we are allowed to apply the conventional
central limit theorem, thereby recovering a diffusion process,
indistinguishable in the long-time limit from those triggered 
by the Wiener fluctuations, with variance per unit of time given by:
\begin{equation}
\sigma^2 = \frac{2\,\beta\, W^2\,T}{\beta - 1} \mbox{ .}
\label{varfree}
\end{equation}
We say \cite{gnekol} that the diffusion process is attracted 
by the basin of Gauss statistics. If $0 < \beta < 1$ the second moment 
of the distribution of Eq.(\ref{dist_dicot}) is not finite, and we 
are therefore forced to use the L\'evy-Khintchine-Gnedenko
 theorem \cite{gnekol} 
(that is, the extension of the central limit theorem
for random variables with no finite variance). 
According to Ref. [17] and to the result of the Appendix, 
this condition yields a L\'evy process with
parameters $b$ and $\alpha$ given by:
\begin{equation}
b = W \,(\beta \, T W)^{\beta} \sin\left(\frac{\pi}{2}\beta\right)\Gamma(1-\beta)
\label{b_free}
\end{equation}
and 
\begin{equation}
\alpha = \beta + 1 \mbox{ .}
\label{alpha}
\end{equation}
In this case we say that the diffusion process is attracted by the
basin of L\'evy statistics.

We see that in the long-time limit 
the diffusion generated by $\xi$ is essentially 
indistinguishable from that generated by the L\'evy fluctuation
$\eta$. What about the influence of an external perturbation on these 
two distinct fluctuations? One might be tempted to make the 
conjecture that both processes result in the same equilibrium 
distribution. However, for this conjecture to be correct it would be 
necessary to apply first the generalized central limit theorem and, 
afterwards, the external perturbation. In general, the result is 
different from what one would obtain reversing this order, 
namely, applying the perturbation first, and the time asymptotic 
analysis next. Thus the naive prediction that both fluctuations 
produce the same equilibrium distribution, in general is wrong.

To substantiate our view, let us consider the \emph{dynamic 
counterpart} of Eq.(\ref{damped}):
\begin{equation}
\frac{d}{dt} v(t) + \lambda v(t) = \xi(t) \mbox{ .}
\label{damped&dyn}
\end{equation}
It is easy to solve Eq.(\ref{damped&dyn}). In a single motion event
we have:
\begin{equation}
v(t)=\frac{\pm W}{\lambda}(1-e^{-\lambda (t-t^{(0)}_{n-1})})
+v(t_{n-1}^{(0)})\,e^{-\lambda (t-t^{(0)}_{n-1})} \mbox{ ,}
\label{solveddamped}
\end{equation}
$$
\mbox{with }\;\;\;
 t_{n-1}^{(0)}<t<t_{n-1}^{(0)}+\tau_n \;\;\; \mbox{ and } \;\;\;
 t_{n-1}^{(0)}=\sum_{k=0}^{n-1}\tau_k \mbox{ .}
$$
Here $v(t)$ is forced to remain in the interval $[-W/\lambda,+W/\lambda]$. 
%so the mean kinetic energy is kept finite.
The computer simulation of this letter is based, as done in 
Ref.\cite{marcoluigi},
on generating, by means of a suitable non-linear
deformation of the random numbers with a uniform distribution in the 
interval $[0,1]$, a sequence of sojourn times   
$\tau_k$ with the density of Eq.(\ref{t_sogg}), and with another
random number generator, equivalent to tossing a coin,
a random sequence of velocity signs. For each trajectory
realization we set the initial condition 
 $v(0)=0$ and the trajectory is observed till to a fixed stop
time, and subsequently recorded in a bin. Each trial was repeated 
$10^4$ times.
In all the numerical calculations behind the results 
illustrated in the enclosed figures, 
we assume that equilibrium is reached after a time $> 20/\lambda$.
The following subsections are devoted to illustrating the results
obtained in three distinct physical conditions.
Note that to illustrate the 
equilibrium distribution shape corresponding to different values of 
$\lambda$ we use a variable $z$, which is the rescaled velocity 
obtained by multiplying the original velocity $v$ by the factor 
$W/\lambda$. Thus the rescaled velocity ranges from $z= -1$ to $z = 1$.

\subsection{Large perturbation}

In the strong damping case, say when $\lambda T \beta > 1$, for any of 
the two accelerating states equilibrium  is reached before the state comes 
to its end. This means that the preferred velocity values will be 
either $W/\lambda$ or $-W/\lambda$. This also means that the equilibrium 
distribution is expected to be $\cup$-shaped, as confirmed by the curve of 
Fig. \ref{formazioneLevy}  corresponding to $\lambda=0.05$. 
This damping-induced 
$\cup$-shaped distribution is generated by system's dynamics regardless of 
whether the corresponding free diffusion is located in the L\'{e}vy or 
the Gauss basin of attraction.

\subsection{Weak perturbation}

\begin{figure}
\begin{picture}(400,270)(0,0)
\epsfysize=6truecm
\epsfxsize=14truecm
\put(0,35){\epsfbox{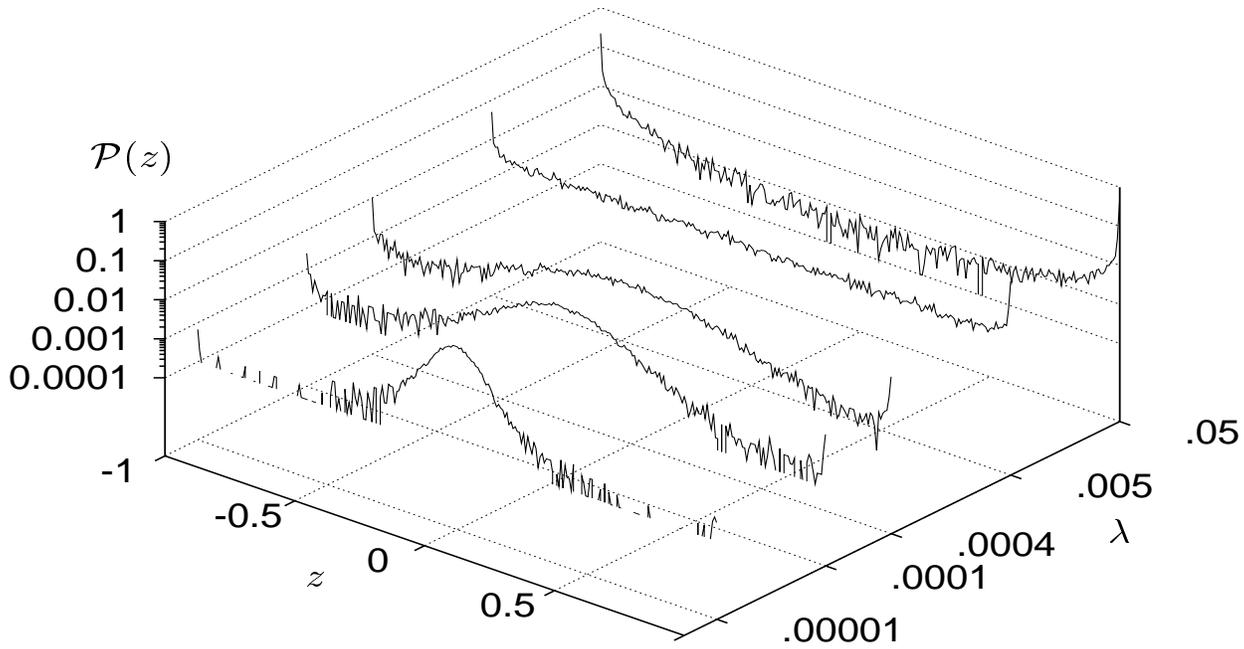}}
\end{picture}
\caption{Equilibrium probability densities for different values of
the friction $\lambda$ at $\beta=0.6$. The L\'evy shape emerges upon
friction decrease.}
\label{formazioneLevy}
\end{figure}

We have seen that in the case of very large friction the 
distribution has a distinct $\cup$-shaped form: This means two peaks 
enclosing an almost empty central region. From Fig. \ref{formazioneLevy}
we see also that 
weakening the perturbation intensity has first the effect of populating 
this empty central region with a uniform distribution (see the curve 
corresponding to $\lambda = 5 \cdot 10^{-3}$). A further decrease of the 
perturbation intensity ($\lambda \leq 4 \cdot 10^{-4}$) makes 
a bell-shaped L\'evy 
distribution of decreasing width and increasing amplitude appear in 
this central part of the equilibrium distribution.

\begin{figure}
\begin{picture}(400,270)(0,0)
\epsfysize=10truecm
\epsfxsize=17truecm
\put(-10,20){\epsfbox{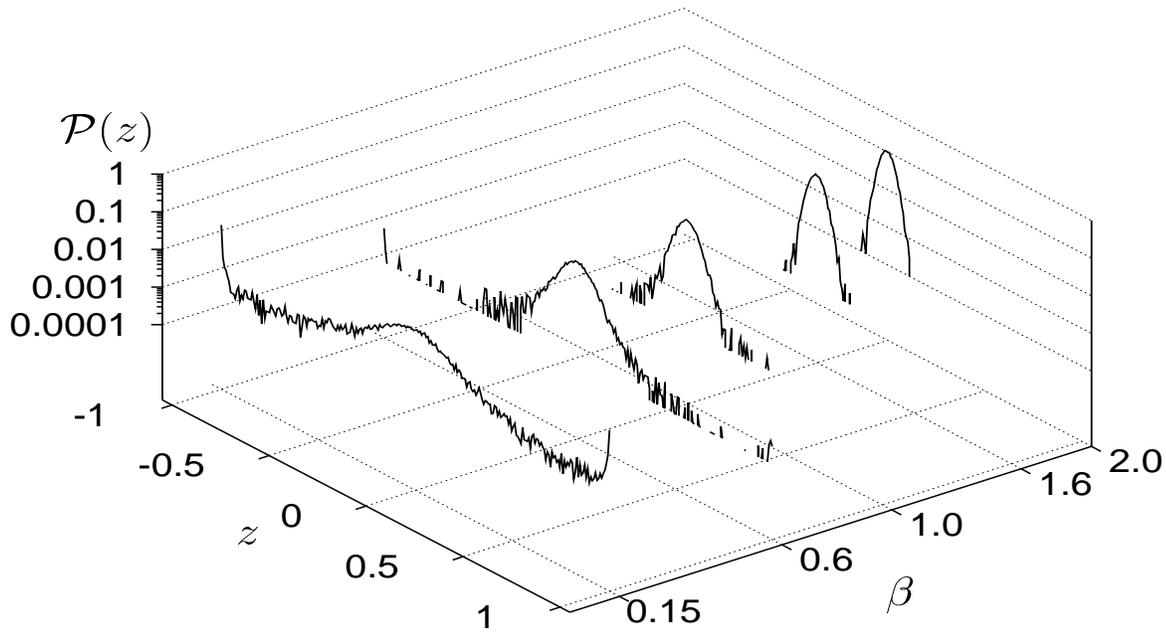}}
\end{picture}
\caption{Equilibrium probability densities for different values of $\beta$ 
at $\lambda = 10^{-5}$. The distribution change shape from L\'evy to
Gauss one when $\beta$ crosses the critical value $\beta=1$.}
\label{lambdacost4}
\end{figure}

Fig. \ref{lambdacost4} is devoted to illustrating the effect that changing 
$\beta$ has on the distribution shape, in a case of weak friction, 
namely, with $\lambda = 10^{-5}$. Note that in the Gauss basin of 
attraction, with $\beta >1$, the distributions are Gaussian functions 
with no algebraic tails, whereas in the L\'{e}vy basin of attraction, 
with $\beta <1$, the equilibrium distribution is markedly characterized 
by the emergence of slow tails. The transition from the one to the 
other condition takes place at $\beta = 1$.

\subsection{Transition from the L\'{e}vy to the Gauss basin of
attraction}

As pointed out in subsection 3 B,
in the weak-friction regime, the statistics of the equilibrium 
distribution are determined by the values of the parameter $\beta$.
According to whether system's dynamics belongs to the Gauss or the the
L\'evy basin of attraction, we are led to adopt different
criteria to determine the distribution width 
using the stochastic analysis, i.e. 
the variance $\sigma^2_\lambda$ or the parameter  $b_\lambda$.
In the former case, $\beta > 1$, the variance $\sigma^2_\lambda$
is obtained using Eq.(\ref{sig_ou}) and Eq.(\ref{varfree}). This yields:
%%%% pictures for two coloumn version
%%%%% here the pictures...
%
%
\onecolumn
\begin{figure}
\begin{picture}(400,270)(0,0)
\epsfysize=10truecm
\epsfxsize=16truecm
\put(-10,10){\epsfbox{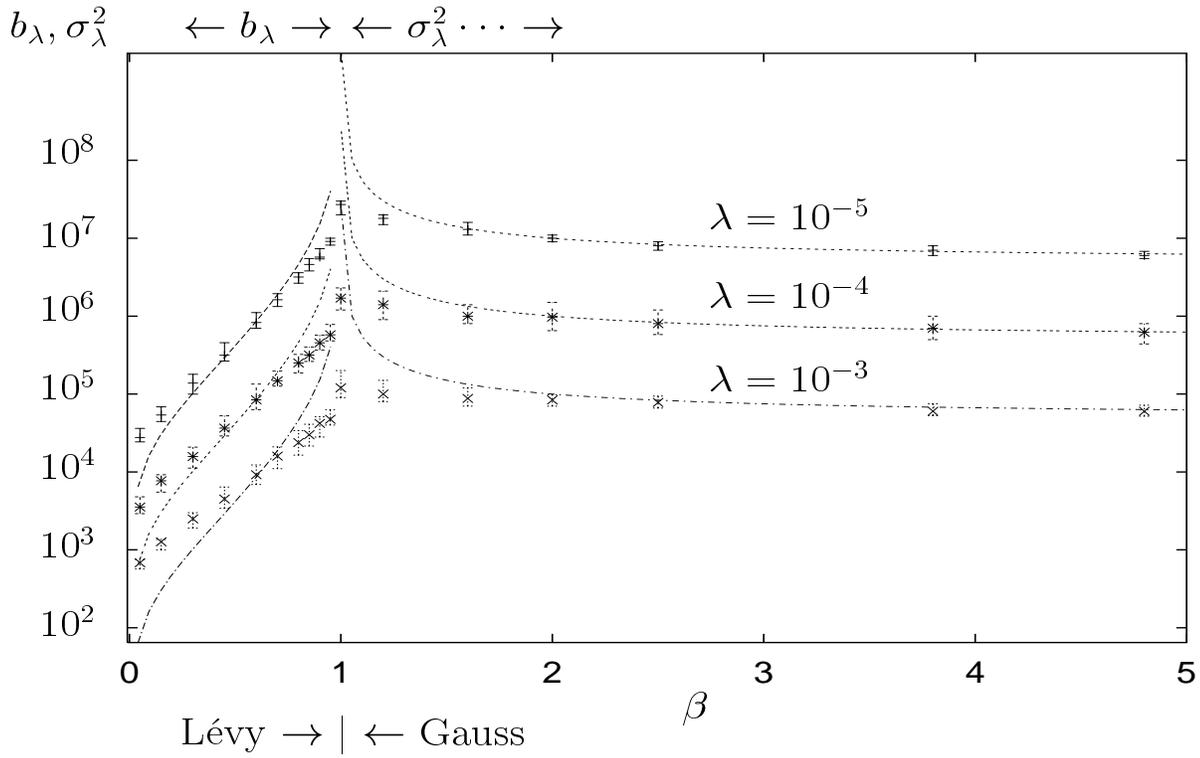}}
\end{picture}
\caption{The distribution widths as a function of $\beta$.
The curves are parametrized by the friction $\lambda$.
For $\beta<1$ the ordinates refer to  $b_\lambda$.
For $\beta>1$ the ordinates refer to $\sigma^2_\lambda$.
The lines illustrate the theoretical prediction
according to the stochastic analysis. 
The points are the numerical results.}
\label{misure_b_beta}
\end{figure}
\begin{figure}
\begin{picture}(400,270)(10,0)
\epsfysize=10truecm
\epsfxsize=17truecm
\put(-20,0){\epsfbox{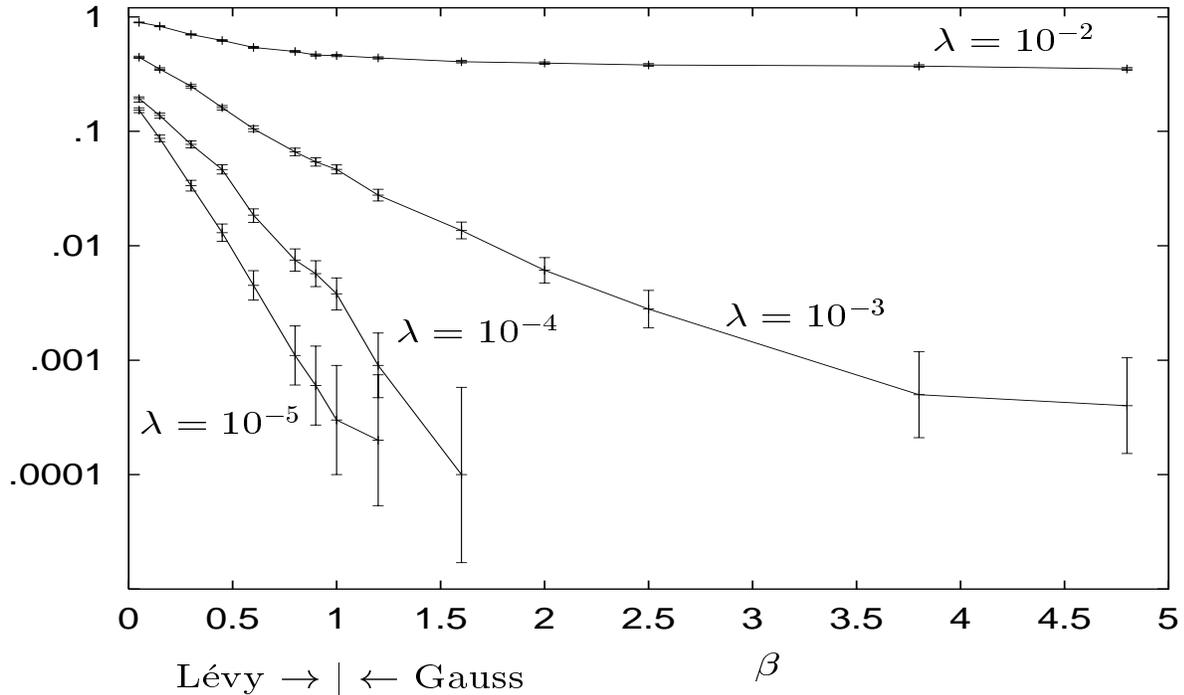}}
\end{picture}
\caption{The population of the region 
$1 > |z| > 0.8$ as a function of $\beta$, for different
values of the friction $\lambda$.
In the small friction condition the population quickly drops to zero  
with $\beta$ moving from $\beta < 1$ to $\beta > 1$.
}
\label{misure_coda}
\end{figure} 
\twocolumn
%%%%
%
\begin{equation}
\sigma^2_\lambda = \frac{\beta\, W^2\,T}{\lambda (\beta - 1)} \mbox{ ,}
\label{sig_sto}
\end{equation}
thereby resulting in a divergence   for $\beta \rightarrow 1^+$.
On the r.h.s. of Fig. \ref{misure_b_beta} we establish a 
comparison between this theoretical prediction and the
corresponding numerical result.
For $\beta >> 1$ we find an excellent 
agreement between theory and numerical simulation.
We note however that the numerical calculations
do not produce any divergence at the transition from the
Gauss to the L\'evy basin of attraction. 
In the latter case, when $\beta < 1$, the expected L\'evy equilibrium 
distribution (see \cite{west}) should be characterized
by the value of the parameter $b_\lambda$ obtained using Eqs.(\ref{b_sw}), 
(\ref{b_free}), (\ref{alpha}). This value is:  
\begin{equation} 
b_\lambda=\frac {W \,(\beta \,
T W)^{\beta}}{\lambda\,(\beta+1)} \sin\left(\frac{\pi}{2}
\beta\right)\Gamma(1-\beta) \mbox{ ,}
\label{b_sto}
\end{equation}
which results again in a divergence at the transition from the L\'evy 
to the Gauss basin of attraction.
In the left side of
Fig.\ref{misure_b_beta} the comparison between theoretical prediction 
and numerical calculation shows a good agreement around
$\beta \simeq 0.5$. 

Although the numerical calculation does not result in any divergence, 
we see that decreasing the friction intensity has the effect of 
improving the agreement between theoretical prediction and numerical
finding. The measured values of the parameters $b_\lambda$ and
$\sigma^2_\lambda$ becomes increasingly larger at the transition from 
one basin of attraction to the other, and the overall behavior becomes
increasingly similar to that of a phase transition.

Fig. \ref{misure_coda} is devoted to 
illustrating the peak contributions to the 
equilibrium distribution. We 
have evaluated numerically the amount of population for
$0.8 \leq |z| \leq 1$. We see that decreasing $\lambda$
has the effect depleting this region if $\beta>1$,
while a significant amount of population 
is left in this region if $\beta < 1$.
This numerical result shows that at extremely small friction values, 
moving from $\beta < 1$ to $\beta > 1$ has the significant effect of making 
the peak intensity drop to zero, even if the phase-transition character 
of the passage from $\beta < 1$ to $\beta > 1$ is 
characterized by rare intense 
fluctuations and, consequently, by a large numerical error.

\section{Concluding remarks}
As its main contribution, this paper sheds light into the difference 
between the conventional stochastic approach and the dynamic approach.
%when studying the Langevin equation (\ref{mainequation}).
Of some relevance is the numerical method adopted. Rather than using 
as stochastic generators intermittent maps we have founded our 
numerical treatment on the random generation of the waiting time 
distribution, a fact that allowed us to settle numerically problems 
that would have implied otherwise  hard numerical difficulties. 
Setting apart the case of strong friction where the stochastic and the 
dynamic method yield strikingly different results, the numerical 
method used made it possible to settle two much more delicate problems:

(i) We have pointed out the residual differences between the two method 
surviving in the extremely weak friction regime.

(ii) We have made accessible to numerical investigation 
the delicate issue of the transition occurring at $\beta = 1$ from the 
Gaussian ($\beta > 1$) to the L\'evy ($\beta < 1$) statistics. 
     
It has to be pointed out that this transition region 
is not yet well understood \cite{bologna} and further research work along the 
lines of this letter might serve the interesting purpose of 
establishing whether in the limiting case of very small friction the 
equilibrium distribution yields
the divergencies predicted by Eq. (\ref{sig_sto}) and Eq.(\ref{b_sto}) 
or a finite
values, as suggested by the recent theoretical analysis
of Ref. \cite{bologna}. We see from Fig. \ref{misure_b_beta}
that the numerical analysis 
yields finite rather than divergent widths at $\beta = 1$, but at 
the moment it is not yet clear if this is due to $T > 0$, as argued in 
Ref. [20], or to  the adoption of not yet sufficiently small values of
$\lambda$.

%(ii) We have shown that can emerge two different kind of statistics 
%according to the integrability condition of the termal bath 
%correlation function.
%(iii) The mean kinetic energy  
\appendix
\section*{}

In this Appendix we show how to determine the parameter $b$ defining the 
width of the diffusion process (\ref{Levyproc}) generated, in the 
long-time limit, by
 the dichotomous variable of Eq.(\ref{dicotfree}). 
To realize this goal we use some fundamental theorems established 
by L\'evy, Kintchine and Gnedenko, whose detailed demonstration can 
be found in Ref.\cite{gnekol}.

The central issue is the assessment of the limit distribution 
of the normalized sum of the independent and identically distributed
random variables $\zeta_1, \zeta_2, \cdots, \zeta_n$:
\begin{equation}
\omega_n = \frac{\zeta_1+\zeta_2+\cdots+\zeta_n}{B_n}-A_n \mbox{ ,}
\label{normsumcano}
\end{equation}
where $A_n$ and $B_n$ are suitable normalization constants.
To establish this limit condition we rest on the following 
three properties. 

First, we define what a stable distribution is all 
about. A distribution $V(x)$ is \emph{stable} if given the arbitrary real 
numbers $a_1>0$, $b_1$, $a_2>0$, $b_2$ there exist the numbers
$a_3>0$ and $b_3$ such that the equality
\begin{equation}
V(a_1 x + b_1) \star V(a_2 x + b_2) = V(a_3 x + b_3)
\label{stability}
\end{equation}
holds. Here $\star$ denotes the convolution product.

Second, we use a theorem 
by Kintchine and L\'evy \cite{gnekol} which establishes that the stable 
distributions are those, and only those, corresponding to the sum of 
Eq.(\ref{normsumcano}) converging to a finite limit as
$n \rightarrow \infty$.

 Third, using another theorem Kintchine and L\'evy \cite{gnekol}, 
we express the stable 
distribution by means of characteristic functions with the form 
$\exp(i\gamma k - c|k|^\alpha )$. The range of $\alpha$ is:
$0<\alpha<2$ , namely 
wider than that compatible with the dynamic treatment of this letter 
that only focuses on the interval: $1<\alpha<2$. Our treatment 
is restricted to the case of symmetric distribution, thereby 
setting  $\gamma = 0$. The same theorem by Kintchine and L\'evy establishes 
that
\footnote{Due to a misprint in Eq.(11) of chap.7 \S 34 of 
Ref.\cite{gnekol}, the 
factor $\alpha$ of Eq. (\ref{b_levkin}) is missing.}:
\begin{equation}
c=-\alpha\,M(\alpha)(c_1+c_2)\cos\left(\frac{\pi}{2}\alpha\right)\;
\;\;\;\;\mbox{$c > 0$,}\;\;\;\;\mbox{$1 < \alpha < 2$ ,}
\label{b_levkin}
\end{equation}
where
\begin{equation}
M(\alpha)=\int_0^\infty (e^{-y}-1+y)\frac{dy}{y^{1+\alpha}}= \\
\frac{\Gamma(2-\alpha)}{\alpha (\alpha-1)} \mbox{ }
\end{equation}
and $c_1$ and $c_2$ are two constants establishing the distribution 
asymptotic properties. 

 Let us consider the distribution function $F(x)$ of the 
random variable $\zeta$. According to a theorem established by 
Gnedenko \cite{gnekol}, the distribution $F(x)$ converges to a 
stable function $V(x)$ in the sense of Eq. (\ref{stability}) if, 
and only if, the following relations
\begin{equation}
F(x)=(c_1 a^\alpha + q_1(x))\frac{1}{|x|^\alpha}\;\;\;\;\mbox{if $x < 0$} 
\label{eqc1}
\end{equation}
and
\begin{equation}
F(x)=1-(c_2 a^\alpha + q_2(x))\frac{1}{x^\alpha}\;\;\;\;\mbox{if $x > 0$} 
\label{eqc2}
\end{equation}
hold. Note that the function $q_1(x)$ and $q_2(x)$ vanish for  $x \rightarrow 
-\infty$ and $x \rightarrow \infty$ , respectively. The constant $a$ 
depends on the choice of 
$B_n$ appearing in Eq.(\ref{normsumcano}). We make the choice 
$B_n = n^{1/\alpha}$ yielding 
$a = 1$. Thus in the the asymptotic limit $|x| \rightarrow \infty$
the distribution $F(x)$ only depends on $c_1$ and $c_2$. Furthermore, 
the earlier choice of a symmetric distribution yields $c_1 = c_2$.

 Note that to realize our purposes we have to adopt the 
non-normalized form:
\begin{equation}
\omega_n^\prime = \Delta v_1+\Delta v_2+\cdots+\Delta v_n \mbox{ .}
\label{notnormsumcano}
\end{equation}
 It is evident that 
this non-normalized form can be related to the normalized form of 
Eq. (\ref{normsumcano}) by setting $\Delta v_k = \zeta_k n^{-1/\alpha}$
and $A_n=0$. 
This is equivalent to 
replacing the parameter $c_1$, responsible, as earlier shown, for the 
asymptotic distribution properties, with $c_1 n$.
Claiming no rigour, we set $n = t/T$, where $T$ is the mean value of the 
interval between two realizations of $\Delta v_k$, so
$n$ is the mean number of realizations of the random variable $\Delta v$. 
Thus we obtain for the stochastic L\'evy process $\exp(-b t |k|^\alpha)$:
\begin{equation}
b=c/T \mbox{ .}
\label{b_per_unit_of_time}
\end{equation}
By identifying 
$F(x)$ with the function distribution of (\ref{dist_dicot}), we obtain:
\begin{equation}
F(x)=1-\frac{1}{2}\frac{(\beta\,T)^{\beta+1}}{(\beta\,T
+x/W)^{\beta+1}}\;\;\;\;\mbox{if $x > 0$.} 
\label{ourF}
\end{equation}
Finally, by applying Eq. (\ref{eqc2}) to Eq. (\ref{ourF}) and using Eqs.
(\ref{b_levkin}) and (\ref{b_per_unit_of_time}),
we obtain the result of Eq. (\ref{b_free}).

\end{document}